\documentclass[12pt]{iopart}
\usepackage{graphicx}
\usepackage{iopams}
\usepackage{amssymb}
\usepackage{multirow}
\usepackage{xcolor}

\newcommand{\ped}[1]{\ensuremath{_{\rm #1}}}

\begin{document}
\title{Multiband superconductivity in high pressure sulfur hydrides}
\author{G.A. Ummarino$^{1,2}$ and A. Bianconi$^{3,4}$}
\address{$^1$ Istituto di Ingegneria e Fisica dei Materiali,
Dipartimento di Scienza Applicata e Tecnologia, Politecnico di
Torino, Corso Duca degli Abruzzi 24, 10129 Torino, Italy.}
\address{$^2$ National Research Nuclear University MEPhI (Moscow Engineering Physics Institute),
Kashira Hwy 31, Moskva 115409, Russia.}
\address{$^3$ RICMASS Rome International Centre Materials Science Superstripes Via dei Sabelli 119A, 00185 Rome, Italy. ORCID: 0000-0001-9795-3913}
\address{$^4$ Institute of Crystallography, National Research Council of Italy, Via Salaria Km 29.300, Monterotondo Stazione, 00015 Rome, Italy.}
\ead{giovanni.ummarino@polito.it}
\ead{antonio.bianconi@ricmass.eu}

\begin{abstract}
The temperature dependence of  the two superconducting gaps in pressurised $H_{3}S$ at $155$ GPa with a critical temperature of $203$ K has been determined  by data analysis of the experimental curve of the upper critical magnetic field as a function of temperature in the framework of the two bands $s$-wave Eliashberg theory, with two different phonon mediated intra-band Cooper pairing channels in a regime of moderate strong couplings with the key role of the pair exchange interaction between the two gaps giving the two non-diagonal terms of the coupling tensor which are missing in the single band s wave Eliashberg theory. The results provide the prediction of the  different temperature dependence of the small and large gaps as a function of temperature which provide evidence of multi-gap superconductivity in $H_{3}S$ .
\end{abstract}
%\pacs{74.70.Xa, 74.20.Fg, 74.25.Kc, 74.20.Mn}
%\keywords{Multibandsuperconductivity, hydrides, Eliashberg equations}
\maketitle
%%%%%%%%%%%%%%%%%%%%%%%%%%%%%%%%%%%%%%%%%%%%%%%%%%%%%%%%%%%%%%%%%%%%%%%%%%%%%%%%%%%%%%%%%%%
\section{Introduction}
The conventional superconducting state of low critical temperature superconductors has been described in 1957 by BCS theory in the dirty limit where multiple bands are reduced to an effective single-band, with a single large isotropic Fermi surface and a large Fermi energy where Cooper pairs are formed by exchange of a low energy phonon. In this classical approximation the critical temperature is controlled by the phonon energy and the electron-phonon interaction. The material dependent critical temperature has been calculated by using the isotropic Eliashberg theory \cite{revcarbi,revmia} with a single electron-phonon spectral function $\alpha^{2}F(\Omega)$ for the average interaction over the Fermi surface. This function in different materials was determined before from inversion of tunnelling data while now it is calculated by density functional theory (DFT). The single band approximations was found to breakdown also for element niobium \cite{Nb} in 1970 but it was considered at that time a mere curiosity.
 In frame-work of the single band approximation the Fermi energy is far from band-edges and therefore the critical temperature does not shows strong anomalous variations as a function of the lattice compressive strain induced by pressure.
On the contrary in the clean limit the multi-band metals with different Fermi surfaces give multi-gap superconductivity where the non diagonal terms in the coupling tensor are determined by an additional fundamental interaction beyond Cooper pairing in quantum matter: the pair exchange between different gaps.
While the hot topic for the majority of scientific community searching for the mechanism of superconductivity in high $T_{c}$ cuprate superconductors was  focusing the search for unconventional pairing mechanisms using a single band approximation, few authors focused on controlling the pair exchange interaction in multigap scenario.
The Bianconi Perali Valletta (BPV) theory predicted that by tuning the perovskite lattice strain or charge density there is an asymmetric dome for the amplification of  the critical temperature with the threshold at the Lifshitz transition for appearing of a new Fermi surface, and  a maximum at the Lifshitz transition for opening a neck in the appearing Fermi surface driven by the Fano-Feshbach resonance due to configuration interactions between large and small gaps in the BEC-BCS regime and  in the BCS regime controlled by the non diagonal terms of the coupling tensor of multigap superconductivity.\cite{BPV}.
Subsequently, in 2001 superconductivity at 40K in magnesium diboride ($MgB_{2}$) was discovered \cite{MgB2} which was described by the two-band Eliashberg theory \cite{Umma3,Umma0} and the BPV theory \cite{inno1,inno2}.
The very large family of iron based superconductors discovered in 2008 \cite{iron}  provided key evidence for for a plurality of scenarios of multiband superconductivity with a number of bands ranging from three to five and the chemical potential tuned near a Fano-Feshbach resonance or shape resonance between superconducting gaps \cite{shape}.
In these last two decades the extensive search for the pairing mechanism in cuprates remained focused on strong correlation and  $d$ wave  symmetry of the order parameter  \cite{Ummad}.
The history of high pressure as a tool to find novel high temperature superconductors begins with the works  \cite{Ashcroft1,Ashcroft2} focusing on pressurised metallic hydrogen and pre-compressed hydrides until the prediction of Duan \cite{Duan} and the experimental discovery \cite{Eremets1} in 2015.
While in 2015 after the discovery it was quickly noticed  the need of  multi-gap superconductivity theory to describe the $203$ K superconductivity in sulfur hydride
\cite{lifshitz} for 8 years superconductivity in high pressure hydrides \cite{Purans} has been thoroughly investigated in the single band approximation \cite{kresin,Pickett}. The experimental hint that multiband superconductivity could be relevant has been first the anomalous pressure dependent isotope effect \cite{band} and later the experimental evidence of the linear trend of the upper critical field ($B_{c2}$) as a function of temperature \cite{Bc2ex}. In this article we analyse the experimental data available
on $B_{c2}$ as a function of temperature using the multiband Eliashberg theory using the simplest possible model but still able to grasp the relevant physics of this system. Since we do not know exactly all the input parameters that the two-band Eliashberg theory needs, we will examine some possible scenarios which will be verified by future experiments. While we were writing this work  Eremets group has shown tunnelling effect measurements \cite{Eremets} on the $H_{3}S$ at a pressure of $155$ GPa with a critical temperature of $203$ K which clearly show two superconducting gaps: a small one at $10$ meV and a large one at $25$ meV at $77$ K. Our work was motivated by these new experimental data \cite{Eremets} which clearly demonstrate that this system is multiband. Our very simple model showed that the experimental data can be easily explained in Eliashberg's two-band $s$-wave theory and that the electron-phonon coupling is of the order of $\simeq 1$ i.e. half of what it had previously been valued \cite{Bc2ex}.

\section{The model}

We will try to explain the experimental data using the simplest possible model which, despite the approximations, still manages to grasp the fundamental physical aspects. Obviously by giving up some approximations but losing in simplicity it will be possible to further improve the agreement with the experimental data.
The electronic structure of the compound $H_{3}S$, can be approximately described by some large Fermi surfaces  and a small appearing Fermi surface \cite{lifshitz}. The relevant bands, following the nomenclature present in the literature, have been named \cite{band} with the numbers $8$, $9$, $10$, $11$. Band $10$ is narrow (about $400$ meV) and is associated with the large gap ($50$ meV) while the other three bands are broad and are associated with the small gap ($20$ meV). In our model the density of states of the band indicated with index $2$ will correspond to band 10 and thus its density of states at the Fermi level ($N_{2}(0)=0.58$ $(cell\cdot eV)^{-1}$) while the band indicated with index $1$ will correspond to bands $8$, $9$ and $11$ with a density of states at the Fermi level ($N_{1}(0)=0.41$ $(cell\cdot eV)^{-1}$) which corresponds to the sum of the contributions of these three bands.
To calculate the gaps and the critical temperature within the two-band Eliashberg equations \cite{Umma1} one has to solve four coupled equations for the gaps $\Delta_{i}(i\omega_{n})$ and the renormalization functions $Z_{i}(i\omega_{n})$, where $i$ is a band index (that ranges between $1$ and $2$) and $\omega_{n}$ are the Matsubara frequencies.
The imaginary-axis equations \cite{Umma3,Umma1,Umma2,Umma4,Umma5} read:
\begin{eqnarray}
&&\omega_{n}Z_{i}(i\omega_{n})=\omega_{n}+ \pi T\sum_{m,j}\Lambda_{ij}(i\omega_{n},i\omega_{m})N^{Z}_{j}(i\omega_{m})+\nonumber\\
&&+\sum_{j}\big[\Gamma^{N}\ped{ij}+\Gamma^{M}\ped{ij}\big]N^{Z}_{j}(i\omega_{n})
\label{eq:EE1}
\end{eqnarray}
\begin{eqnarray}
&&Z_{i}(i\omega_{n})\Delta_{i}(i\omega_{n})=\pi
T\sum_{m,j}\big[\Lambda_{ij}(i\omega_{n},i\omega_{m})-\mu^{*}_{ij}(\omega_{c})\big]\times\nonumber\\
&&\times\Theta(\omega_{c}-|\omega_{m}|)N^{\Delta}_{j}(i\omega_{m})
+\sum_{j}[\Gamma^{N}\ped{ij}-\Gamma^{M}\ped{ij}]N^{\Delta}_{j}(i\omega_{n})\phantom{aaaaaa}
 \label{eq:EE2}
\end{eqnarray}
In these equations we have defined
\[\Lambda_{ij}(i\omega_{n},i\omega_{m})=2
\int_{0}^{+\infty}d\Omega \Omega
\alpha^{2}_{ij}F(\Omega)/[(\omega_{n}-\omega_{m})^{2}+\Omega^{2}] \]
and
$N^{\Delta}_{j}(i\omega_{m})=\Delta_{j}(i\omega_{m})/
{\sqrt{\omega^{2}_{m}+\Delta^{2}_{j}(i\omega_{m})}}$ and
$N^{Z}_{j}(i\omega_{m})=\omega_{m}/{\sqrt{\omega^{2}_{m}+\Delta^{2}_{j}(i\omega_{m})}}$.
The parameters $\Gamma^{N}\ped{ij}$ and $\Gamma^{M}\ped{ij}$ are the scattering rates from non-magnetic and magnetic impurities while $\Theta$ is the Heaviside function and $\omega_{c}$ is a cutoff
energy.  The quantities $\mu^{*}_{ij}(\omega\ped{c})$ are the elements of the $2\times 2$
Coulomb pseudopotential matrix.
The electron-phonon coupling constants are defined as $\lambda_{ij}=2\int_{0}^{+\infty}d\Omega\frac{\alpha^{2}_{ij}F(\Omega)}{\Omega}$.
In the more general situation the solution of equations \ref{eq:EE1} and \ref{eq:EE2} requires a large number of input parameters (four functions and twelve constants). We have to introduce in the equations four electron-phonon spectral functions $\alpha^{2}_{ij}F(\Omega)$, four elements of the Coulomb pseudopotential matrix $\mu_{ij}^{*}(\omega_{c})$,
four nonmagnetic $\Gamma^{N}\ped{ij}$ and four paramagnetic $\Gamma^{M}\ped{ij}=0$ impurity-scattering rates.
However, some of these parameters can be extracted from experiments and some can be fixed by suitable approximations.
In particular, we put $\Gamma^{N}\ped{ij}=0$ because, when the order parameter is in $s$ wave, the intraband component of impurity scattering rate has not influence on superconductive properties (Anderson theorem) while the interband contribution usually has a negligible effect \cite{UmmaMgB2}. This parameter ($\Gamma^{N}\ped{ij}$) is relevant just for a strongly disordered non $s$-wave superconductors but that's not the case. The same can be done for the scattering rate from magnetic impurities that are absent so we put $\Gamma^{M}\ped{ij}=0$.
We assume that the shape of the electron-phonon functions is the same and we rescale them by changing the parameter $\lambda_{ij}$ i.e. we write $\alpha^{2}_{ij}F(\Omega)=\lambda_{ij}\alpha^{2}F(\Omega)$ and the function $\alpha^{2}F(\Omega)$, normalized to have electron phonon couplig constant equal to one, is taken from literature \cite{a2f} and was calculated for a pressure value very close to the experimental one.
The values of interband coupling are not indipendent because $\lambda_{ij}=\frac{N_{j}(0)}{N_{i}(0)}\lambda_{ji}$ as the interband Coulomb pseudopotential $\mu^{*}_{ij}=\frac{N_{j}(0)}{N_{i}(0)}\mu^{*}_{ji}$.
$N_{i}(0)$ is the density of states at the Fermi level for the band $i$.
We know the values of $N_{1}(0)$ and $N_{2}(0)$ so, in this way, we have six free parameters: 3 $\lambda_{ij}$ and 3 $\mu^{*}_{ij}$.
For reducing the number of free parameters we suppose that the Coulomb pseudopotential $\mu^{*}_{ij}$ is zero.
This choice was made solely to reduce the number of free parameters and in any case does not substantially change the physical picture of the system. In this way
we will determine the minimal values of the electron-phonon coupling constant because if the Coulomb pseudopotential is
different from zero the coupling constants will be slightly greater. The Coulomb pseudopotential usually is a relevant parameter just for the fullerenes \cite{Ummafull}.
At the end we have three free parameters: $\lambda_{11}$, $\lambda_{21}$ and $\lambda_{22}$.
The strategy is to reproduce (if it is possible) exactly the experimental values of the small gap at $T=77$ K ($\Delta_{2ex}=10$ meV)\cite{Eremets} and the experimental critical temperature $T_{cex}=203$ K.
In this way we fix the three free parameters and after we can calculate other physical observables and compare them with experiments.
Finally we use for obtaining the numerical solution of Eliashberg equations a cut-off energy $\omega_{c}=1090$ meV and a maximum quasiparticle energy $\omega_{max}=1100$ meV.
We will examine three cases that roughly exhaust all possible cases: first case with loosely coupled bands, second case with intermediated coupled bands and thirthy case with strongly coupled bands.
Of course in all cases examined the calculated critical temperature exactly reproduce the experimental one.
\section{Results}
In the case of loosely coupled bands the electron-boson coupling-constant matrix $\lambda_{ij}$ becomes:
\begin{equation}
\vspace{2mm} %
\lambda_{ij}= \left (
\begin{array}{ccc}
  \lambda_{11}=0.7250              &         \lambda_{12}=0.0071                            \\
  \lambda_{21}=0.0050              &         \lambda_{22}=1.4931           \\
\end{array}
\right ) \label{eq:matrix}
\end{equation}
to produce a $\lambda_{tot}=((\lambda_{11}+\lambda_{12})N_{1}(0)+(\lambda_{21}+\lambda_{22})N_{2}(0))./(N_{1}(0)+N_{2}(0))=1.1808$ that represents the average of electron-phonon coupling.
In figure 1 the self energy functions $\Delta(\omega)$ and $Z(\omega)$ obtained by
solving the Eliashberg equations in the real axis formulation, at $T=10$ K, are shown while in the upper and lower insets are shown respectively the electron-phonon spectral function $\alpha^{2}F(\Omega)$ and
the superconductive density of states, always at $T=10$ K.
The value of the small gap at $T=10$ K is equal to $14.8$ meV and becomes equal to 10 meV at $T=77$ K that is exactly the measured value.
The big gap at $T=10$ K is equal to $39.6$ meV and becomes equal to 38.7 meV at $T=77$ K, as it is shown in figure 2, a value grater than the experimental value \cite{Eremets} of $25$ meV .
In the case where the bands have an intermediate coupling, the electron-boson coupling-constant matrix $\lambda_{ij}$ is:
\begin{equation}
\vspace{2mm} %
\lambda_{ij}= \left (
\begin{array}{ccc}
  \lambda_{11}=0.5000             &         \lambda_{12}=0.0651                             \\
  \lambda_{21}=0.0460             &         \lambda_{22}=1.5327           \\
\end{array}
\right ) \label{eq:matrix}
\end{equation}
to produce a $\lambda_{tot}=1.1589$. Also in this case the value of the small gap at $T=77$ K is exactly reproduced while at $T=10$ K the values of the two gaps are $12.8$ meV and $42.8$ meV respectively.
In the case where the bands are strongly coupled the electron-boson coupling-constant matrix $\lambda_{ij}$ is:
\begin{equation}
\vspace{2mm} %
\lambda_{ij}= \left (
\begin{array}{ccc}
  \lambda_{11}=0.0050              &         \lambda_{12}=0.1910                            \\
  \lambda_{21}=0.1350              &         \lambda_{22}=1.5849           \\
\end{array}
\right ) \label{eq:matrix}
\end{equation}

to produce a $\lambda_{tot}=1.0888$. Also in the last case the value of the small gap at $T=77$ K is exactly reproduced while at $T=10$ K the values of the two gaps are $10.3$ meV and $43.1$ meV respectively.
In this case there is very little difference between the small gap values at $10$ K and $77$ K because the bands are strongly coupled and the temperature behavior is completely different as it is possible to see in Figure 2.
We can observe that in all cases we are in a regime of moderate strong coupling ($\lambda_{tot}=1.1808$, $\lambda_{tot}=1.1589$ and $\lambda_{tot}=1.0888$) less than lead ($\lambda_{tot}=1.5500$).
Of course this model is very simple and, for example, to improve the agreement with the experimental data (large gap) one should consider that the density of states around the Fermi level is not constant but would make the model more complicated \cite{Ummachi} and add nothing really essential to the physical explanation of the system.
At this point, to decide which of the three cases examined is more plausible, we need to examine other experimental data such as the temperature dependence of the upper critical magnetic field.
%%%%%%%%%%%%%%%%%%%%%%%%%%%%%%%%%%%%%
%%% UPPER CRITICAL MAGNETIC FIELD
The multiband Eliashberg model developed above can also be used to
explain the experimental temperature dependence of the upper critical magnetic field.
The experimenta data \cite{Bc2ex} show that $B_{c2}(T)$ displays a linear dependence on temperature over an extended range as found in a strong coupling one band superconductor or in a multiband superconductor.
Now we check if our model is able to explain the experimental data. We will calculate the temperature behaviour of the upper critical magnetic field
in the three cases examined and we will see if these experimental data will allow us to decide which of the three cases best describes this system.
For the sake of completeness, we give here the linearized gap equations in the
presence of a magnetic field with non-magnetic impurity scattering \cite{UmmaBc2,carbiBc2}. In the following, $v_{Fj}$
is the Fermi velocity of band $j$, and $B_{c2}$ is the upper critical magnetic field:
\begin{eqnarray}
\omega_{n}Z_{i}(i\omega_{n})\hspace{-2mm}&=&\hspace{-2mm}\omega_{n}+\pi
T\sum_{m,j}[\Lambda_{ij}(i\omega_{n}-i\omega_{m})+\delta_{n,m}\frac{\Gamma^{N}\ped{ij}}{\pi T}]\mathrm{sign}(\omega_{m})\nonumber\\
Z_{i}(i\omega_{n})\Delta_{i}(i\omega_{n})\hspace{-2mm}&=&\hspace{-2mm}\pi
T\sum_{m,j}\{[\Lambda_{ij}(i\omega_{n}-i\omega_{m})-\mu^{*}_{ij}(\omega_{c})]\cdot \nonumber\\
& &
\hspace{-2mm}\cdot\Theta(\omega_{c}-|\omega_{m}|)+\delta_{n,m}\frac{\Gamma^{N}\ped{ij}}{\pi T}\}\chi_{j}(i\omega_{m})Z_{j}(i\omega_{m})\Delta_{j}(i\omega_{m})\nonumber
%\label{eq:EE2}
\end{eqnarray}
\begin{eqnarray}
\chi_{j}(i\omega_{m})\hspace{-2mm}&=&\hspace{-2mm}(2/\sqrt{\beta_{j}})\int^{+\infty}_{0}dq\exp(-q^{2})\cdot\nonumber\\
& & \hspace{-2mm}\cdot
\tan^{-1}[\frac{q\sqrt{\beta_{j}}}{|\omega_{m}Z_{j}(i\omega_{m})|+i\mu_{B}B_{c2}\mathrm{sign}(\omega_{m})}]\nonumber
%\label{eq:EE2}
\end{eqnarray}
with $\beta_{j}=\pi B_{c2} v_{Fj}^{2}/(2\Phi_{0})$.
We know that the ratio of the Fermi velocities of the two bands \cite{band} is approximately equal to $0.13$
so we assume that $v_{F2}=v_{F1}\cdot 0.13$. In this way, using the previously used electron-phonon coupling constants, we will fix the Fermi velocity relative to band $2$ in order to obtain the best fit of the experimental data.
We find in the first case (weak coupling between the two bands) $v_{F2}=8.455$ $10^{5} m/s$ and consequently $v_{F1}=6.504$ $10^{6} m/s$.
The value of upper critical magnetic field at very low temperature is $B_{c2}(T=1 K)=98.3$ $T$.
We proceed in the same way in the second case where the bands have an intermediated coupling between them.
Now we find $v_{F2}=8.511$ $10^{5} m/s$ and $v_{F1}=6.547$ $10^{6} m/s$ with $B_{c2}(T=1 K)=106.2$ $T$.
For the last case where the bands are strongly coupled we find $v_{F2}=8.595$ $10^{5} m/s$ and $v_{F1}=6.612$ $10^{6} m/s$ with $B_{c2}(T=1 K)=99.7$ $T$.
In figure 3 is shown the theoretical curves relative to three cases compared with experimental data \cite{Bc2ex}.
It is clearly seen that in all cases the experimental measurements can be reproduced well. In order to decide which of the three cases best describes the physics of the system, it would be necessary to make tunnelling measurements to determine the trend of the two gaps as a function of the temperature which, as we can see in figure 2, is profoundly different in the three cases.
Most likely it is possible to reproduce the upper critical field as a function of temperature also with the one-band Eliashberg theory but with double the electron phonon coupling value \cite{Bc2ex}.
We have not produced this calculation solely because the experimental data of tunneling presented by Emerets \cite{Eremets} clearly show two distinct values of the gaps.
%%%%%%%%%%%%%%%%%%%%%%%%%%%%%%%%%%%%%%%%%%%%
\section{Conclusions}
We have been able to shed light on the quantum mechanism giving the high critical temperature of pressurised $H_3S$ driven by unconventional anisotropic electron-phonon interaction in different portions of multiple Fermi surfaces and pair exchange interaction, like the Majorana exchange force in nuclear heterogeneous matter made of multiple components \cite{vittorini} and the coexistence of weak and strong electron-phonon interaction. The critical temperature, the small gap value and the behaviour of the upper critical field in function of temperature is well reproduced in the framework of two bands s-wave Eliashberg theory with Cooper pairing in a regime of moderate strong coupling and a small essential pair exchange interaction missing in the standard single band model. We predict a different temperature dependence of the large and small gap in the intermediate two-gap regime proposed for pressurised sulfur hydride. To establish which of the three discussed regimes is the correct one, it will be necessary to get further experimental data on the temperature dependence of the two gaps by using tunnelling measurements.
\section{ACKNOWLEDGMENTS}
G.A. Ummarino acknowledges partial support from the MEPhI.\\
%%%%%%%%%%%%%%%%%%%%%%%%
\\
%BILBLIOGRAFIA

%%%%%%%%%%%%%%%%%%%%%%%%
%
\newpage
\begin{figure}
\begin{center}
\includegraphics[keepaspectratio, width=0.9\columnwidth]{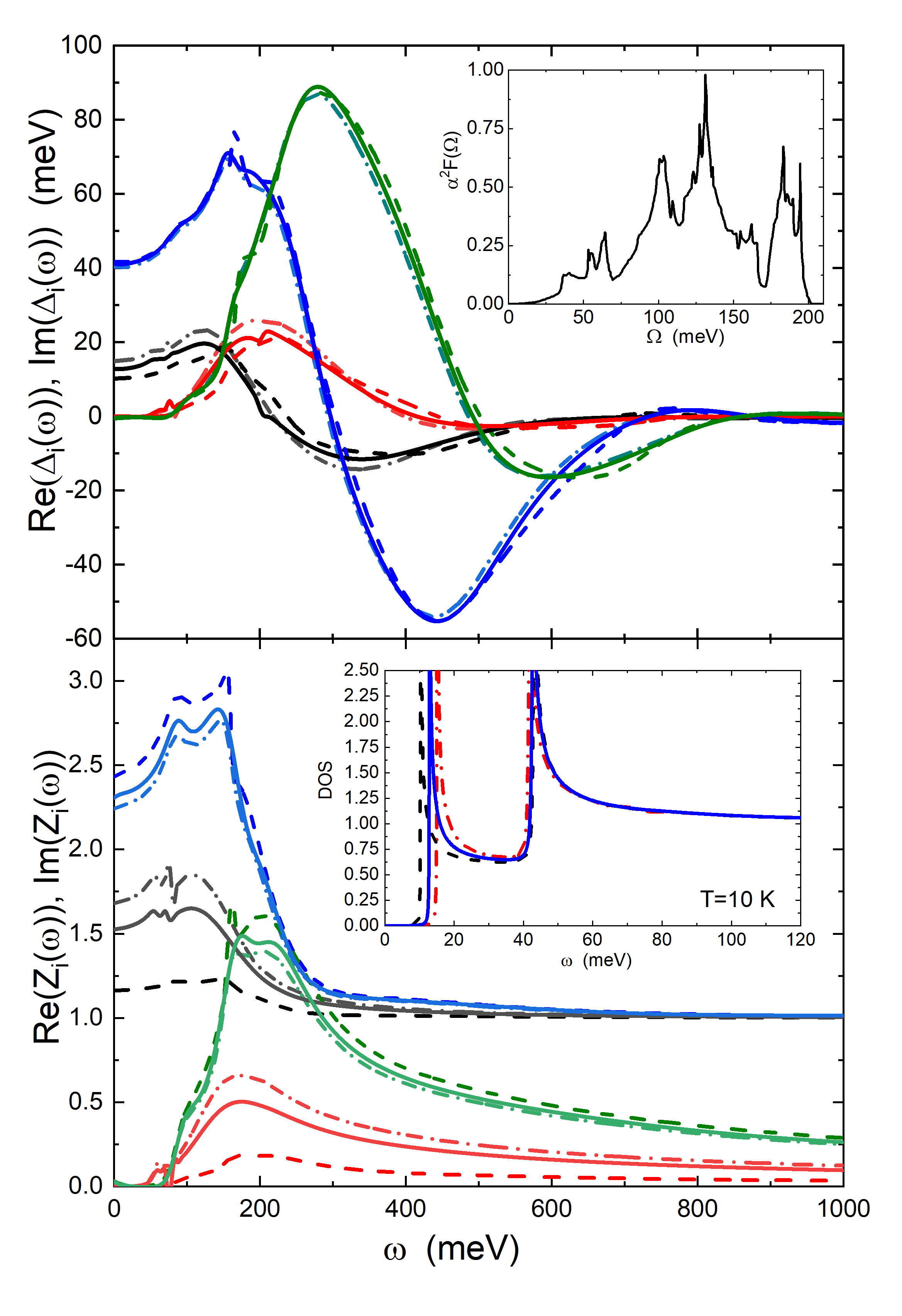}
\vspace{-5mm} \caption{(Color online)
The real and imaginary parts of the self energy functions $\Delta(\omega)$ and $Z(\omega)$ obtained by
solving the Eliasberg equations, at $T=10$ K, in the real axis formulation, in the three cases examined (weak bands coupling dashed dotted lines, intermediate bands coupling solid lines and strong bands coupling dashed lines), at $T=10$ K, are shown as functions of energy $\omega$ in the upper ($\Delta_{i}(\omega)$) and lower ($Z_{i}(\omega)$) panels.
In the inset of the upper panel the electron-phonon spectral function $\alpha^{2}F(\Omega)$ is shown while in the inset of lower panel the superconductive densities of states at $T=10$ K, are shown (weak case red dashed dotted line, intermediate case dark blue solid line and strong case dashed black line).
 }\label{Figure1}
\end{center}
\end{figure}

\newpage
\begin{figure}
\begin{center}
\includegraphics[keepaspectratio, width=\columnwidth]{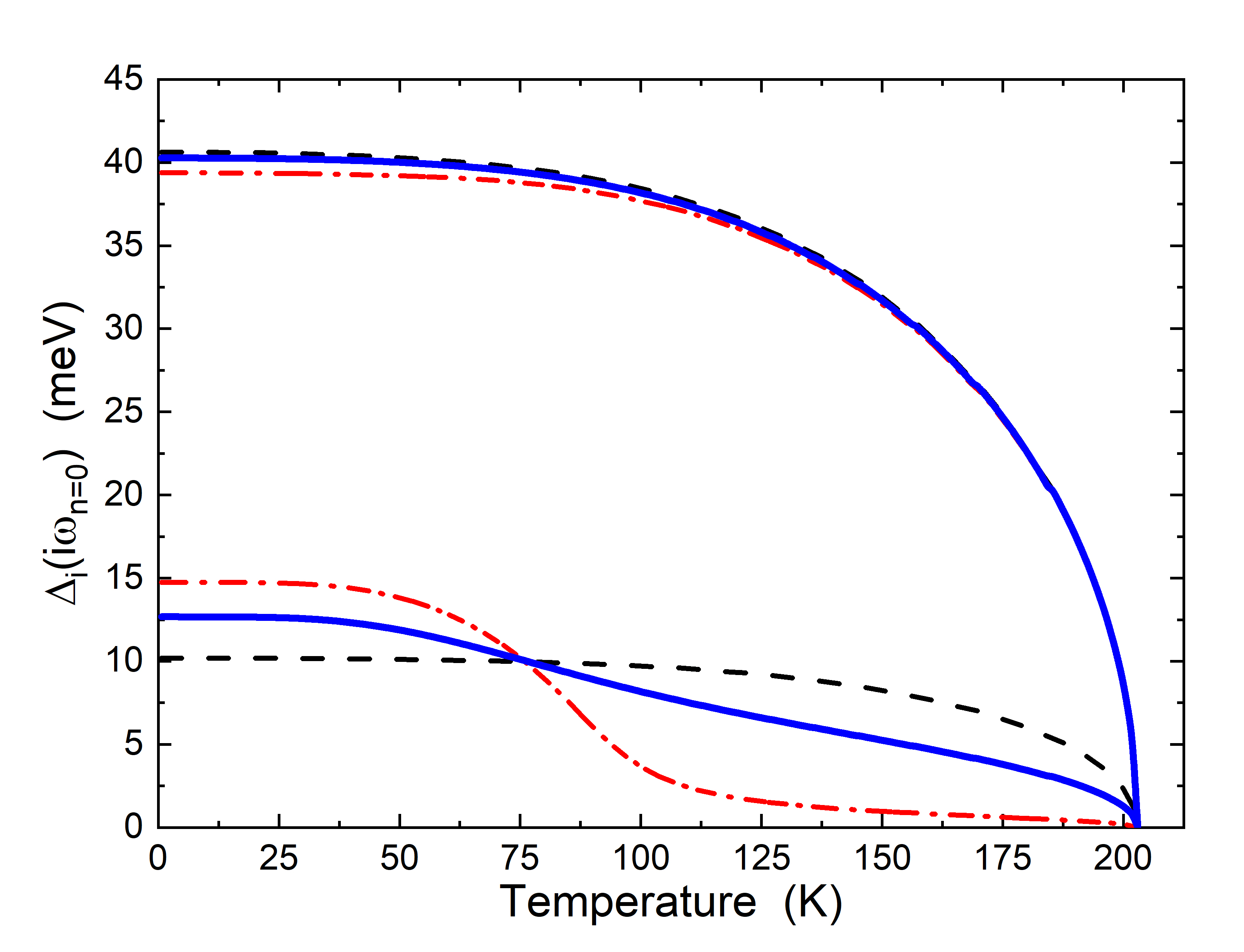}
\vspace{-5mm} \caption{(Color online)
Calculated temperature dependence of the two gaps is shown (weak case red dashed dotted line, intermediate case dark blue solid line and strong case dashed black line).
 }\label{Figure2}
\end{center}
\end{figure}

\newpage
\begin{figure}
\begin{center}
\includegraphics[keepaspectratio, width=\columnwidth]{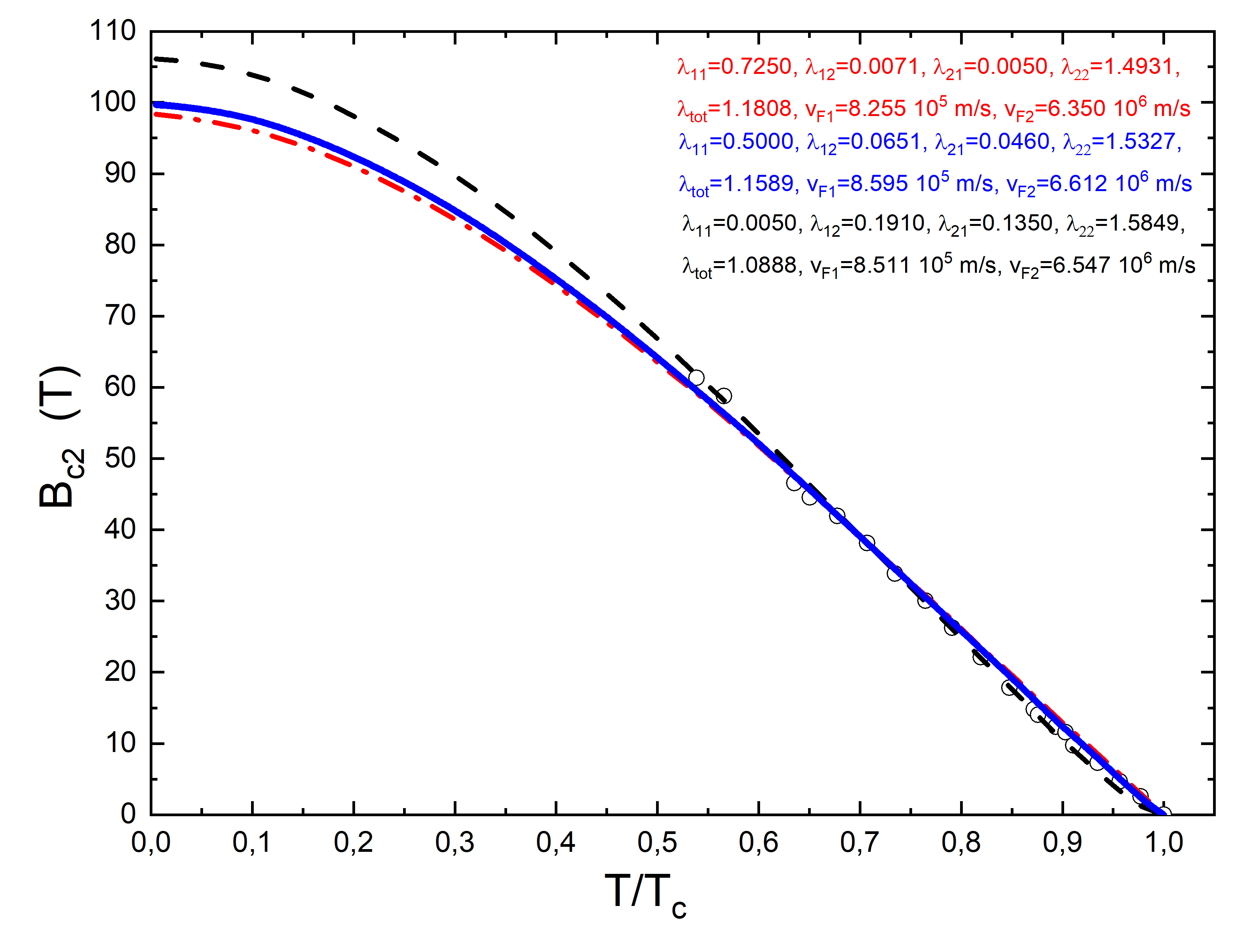}
\vspace{-5mm} \caption{(Color online)
 Experimental temperature dependence of the upper critical field in $H_{3}S$(open circles) from literature \cite{Bc2ex} and the theoretical curves (weak case red dashed dotted line, intermediate case dark blue solid line and strong case dashed black line) obtained by solving the Eliasberg equations for the upper magnetic field with the input parameters of the three cases.
 }\label{Figure3}
\end{center}
\end{figure}

\end{document}